\documentclass[pra, twocolumn, showpacs, longbibliography]{revtex4-1}
\usepackage{epsfig}
\usepackage{amsmath}
\usepackage{dcolumn}
\usepackage{bm}
\usepackage{graphicx}
\usepackage{wrapfig}
\usepackage{graphicx}
\usepackage{epstopdf}
\usepackage{epsfig}
\usepackage{calrsfs}
\usepackage{IEEEtrantools}
\usepackage{color}

\usepackage{amsmath}    
\usepackage{graphicx}   
\usepackage{verbatim}   
\usepackage{color}      
\usepackage{subfigure}  
\usepackage{hyperref}   

\newcommand{\nn}{\nonumber}

\newcommand{\p}{\partial}

\newcommand{\Rbg}{Rabi-Bogoliubov}
\newcommand{\Bg}{Bogoliubov}
\newcommand{\BH}{Bose-Hubbard}

\newcommand{\wh}{\widehat}

\begin{document}



\title{  {Damping-free collective oscillations of a driven two--component
Bose gas in optical lattices}}
\author{Gavriil Shchedrin, Daniel Jaschke, and Lincoln D. Carr}
\affiliation{Colorado School of Mines, Golden, Colorado 80401, USA}

\begin{abstract}
We explore quantum many-body physics of a driven Bose-Einstein condensate in optical lattices.  {The laser field induces a gap in the generalized Bogoliubov spectrum proportional to the effective Rabi frequency. The lowest lying modes in a driven condensate
are characterized by zero group velocity and non-zero current. Thus, the laser field induces roton modes, which carry interaction in a driven condensate. We show that collective excitations below the energy of the laser-induced gap remain undamped, while above the gap they are characterized by a significantly suppressed Landau damping rate.}
\end{abstract}


\maketitle

Multicomponent Bose-Einstein condensates (BECs) of ultracold gases \cite {stamper2013spinor, kawaguchi2012spinor, ueda2010fundamentals,pethick2008bose}
are a superb system for exploring quantum many-body physics and emergent
phenomena in a  {well-controlled macroscopic quantum system}.
The spinor BEC led to a major advances that include observation of Dirac monopoles \cite{ray2014observation}, exotic magnetism  \cite{stamper2013spinor}, spin Hall effect
\cite{li2014chiral}, spontaneous symmetry breaking \cite{sadler2006spontaneous}, coherent spinor dynamics \cite{chang2005coherent},
dynamic stabilization \cite{hoang2013dynamic}, vortex formation \cite{schweikhard2004vortex},
quantum spin mixing \cite{law1998quantum},  spin domain wall formation \cite{stenger1998spin}, and realization of topological states \cite{choi2012observation, williams1999preparing}. The observation of the low-lying collective modes
of a condensate revealed its dynamics and unique spectral signatures \cite{jin1996collective, mewes1996collective}.  Moreover, collective modes in a two-component condensate were instrumental in the examination of the crossover from a BEC to Bardeen-Cooper-Schrieffer (BCS) superfluid regime
{of ultracold Fermi gases} \cite{chen2005bcs, bourdel2004experimental, bartenstein2004collective}. {At low temperatures, the decay of collective modes, caused by the coupling between them, is described by the Landau damping process
\cite{pixley2015damping}}.  {In this Letter, we show that the laser field induces a gap in
an otherwise gapless Bogoliubov spectrum, which leads to the existence of roton modes in a driven condensate. We show that the laser-induced gap in the spectrum of elementary excitation protects the low-lying collective modes from Landau damping.
Above the energy of the gap, the damping is dominated by the laser-induced roton modes
and is considerably suppressed compared to the phonon-mediated damping found in a field-free condensate.}

 \begin{figure}[h!]
\centering
\hspace{-1.0cm}
\subfigure[]
{\includegraphics[width=0.75\columnwidth]{{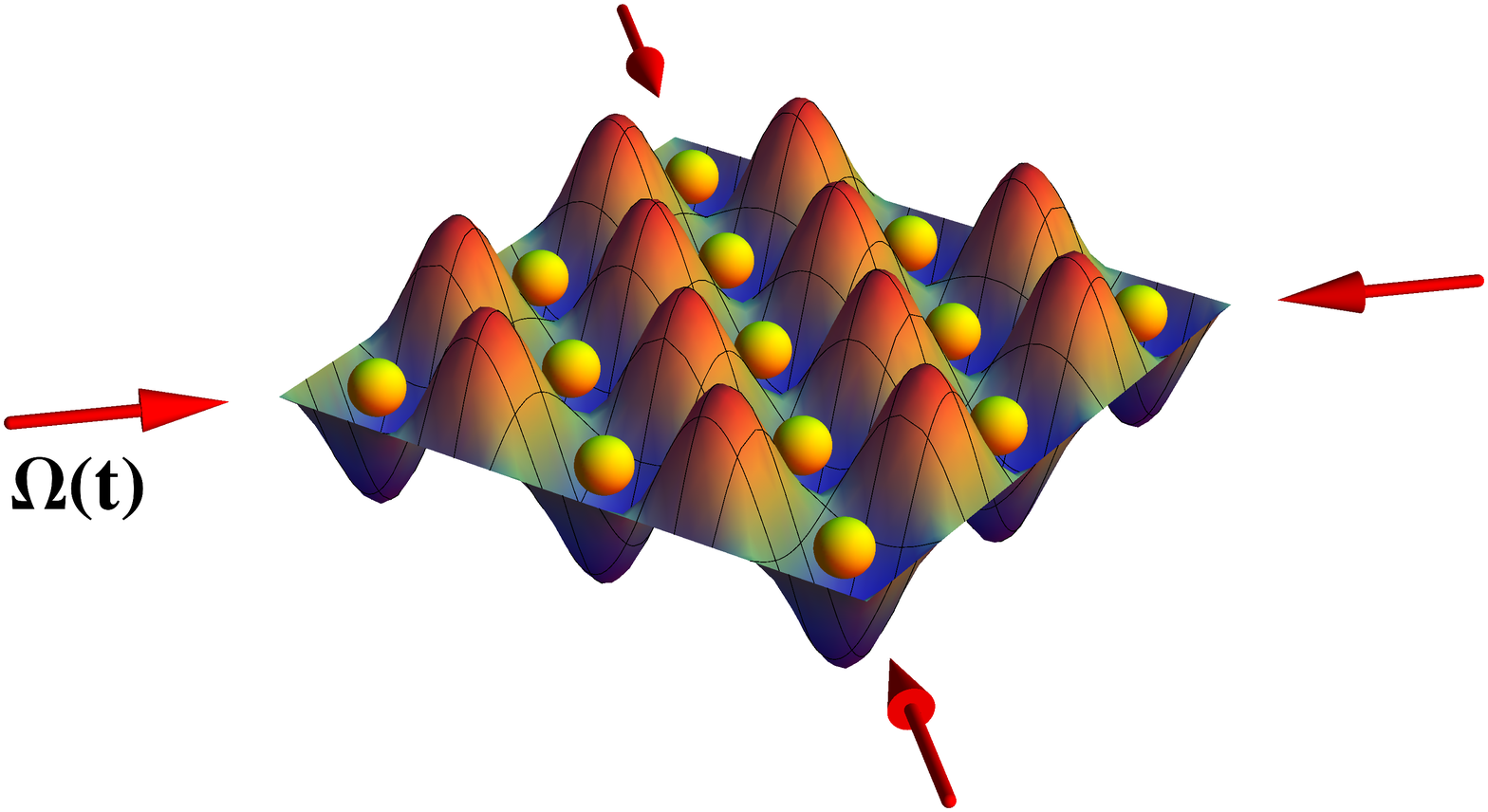}}}
\hspace{-0.5cm}
\subfigure[]
{\includegraphics[width=0.3\columnwidth]{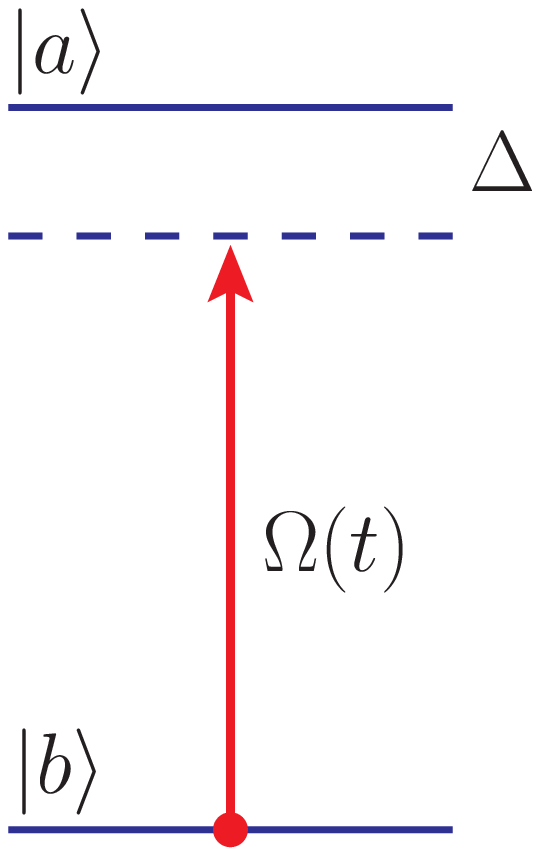}}
\caption{ {\it Bose-Einstein condensate  in an optical lattice driven by a laser field}  (a) Bose-Einstein condensate (yellow spheres) {confined in an optical lattice}, formed by counter-propagating waves (red arrows) and driven from the ground state by the laser field  $\Omega(t)$.  (b) Two-component Bose-Einstein condensate is modeled by a set of two-level systems, driven from the ground state $|b\rangle$ to the excited state
$|a\rangle$ by the laser field $\Omega(t)$. The laser field is characterized by the Rabi  frequency $\Omega_{R}$ and detuning $\Delta$. Initially, the condensate is prepared in the ground state $|b\rangle$.}
\label{fig1}
\end{figure}

We describe a weakly interacting two-component BEC confined in an optical lattice
(see Fig.{\ref{fig1}.(a)) by a driven {\BH} Hamiltonian \cite{pethick2008bose, ueda2010fundamentals}. For a driven two-component BEC we obtain the exact results for the elementary amplitudes and find a set of exact symmetries that are inherent among them. Based on the obtained energy spectrum and amplitudes, we explore the near-equilibrium dynamics of a condensate and calculate Landau decay rate of the collective modes.
The microwave field that drives the condensate from the ground state to the first excited state operates in a single mode regime \cite{jaksch2005cold, sorensen2001many, goldstein1, Abad1, Lellouch1}, and is characterized by the Rabi frequency and detuning (see Fig. {\ref{fig1}.(b)).   { We find that the applied laser field creates a gap in the energy spectrum that dramatically modifies the interaction in a driven condensate. In a scalar BEC, coupling between the collective modes, carried by the phonons,  leads to the absorption (emission) of the collective modes described by Landau damping \cite{landau1946vibrations} {(Beliaev damping \cite{beliaev1958application})}. In contrast, the lowest lying elementary excitations in a driven BEC have zero group velocity and non-zero current. Thus, the interaction in a driven condensate is carried by the laser-induced roton modes. The laser-induced gap in the spectrum of elementary excitation ensures zero Landau damping of the collective modes lying below energy of the gap. Above the gap, it is proportional to the density of the laser-induced roton mode and is considerably slowed down compared to a field-free scalar condensate. }

 {The suppression of the collective modes was previously reported for a number of physical systems. This includes observation of suppressed Landau damping in the Bose-Fermi superfluid mixture \cite{ferrier2014mixture, PhysRevLett.113.265304},
reduced decay rate of the collective excitation in fermionic polar molecules confined in optical lattices due to the quantum Zeno mechanism
\cite{yan2013observation},
and prediction of absence of damping for quasi 2D dipolar Bose gas at zero temperature
\cite{PhysRevA.88.031604,PhysRevA.88.063638, PhysRevA.90.043617}.
Here we report dramatic modification of the spectral signatures and decay rate of the collective modes for a driven two-component Bose gas. The undamped collective modes of a driven multicomponent condensate, combined with its extremely long coherence time, will allow one to observe long-lived internal dynamics of nonlinear macroscopic
phenomena, such as quantum vortices, quantum turbulence, and solitons \cite{ueda2010fundamentals}.}

\begin{figure}[t!]\label{f1}
\begin{center}
\includegraphics[angle=0, width=1.0\columnwidth]
{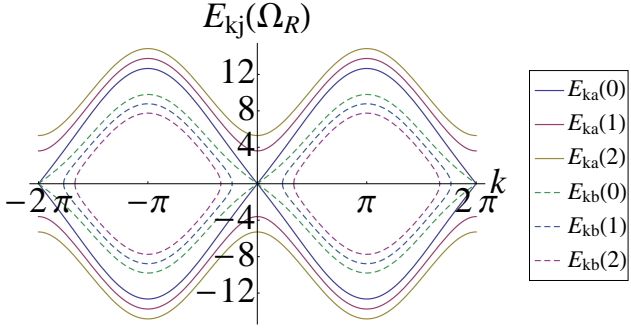}
\end{center}
 \caption{{\it{\Rbg} spectrum for a driven two-component Bose-Einstein condensate in an optical lattice}. The new dispersion law consists of
 two branches,$E_{ka}(\Omega_{R})$ and  $E_{kb}(\Omega_{R})$, which are separated by the standard  {\Bg} spectrum. The first branch of the new {\Rbg} spectrum that lies above the  {\Bg} spectrum is characterized by the gap and has characteristic quadratic dispersion relation for small momenta $k$ in
the units of the lattice constant $a_{L}$. The second branch of the spectrum has a real solution only for certain values of momenta, while for the rest of the momenta, has only purely imaginary solutions.}
\label{spectrum1}
\end{figure}

We start with the Hamiltonian of the two-component BEC in an optical lattice,
\cite{micheli2003many, goldstein2000eliminating, PhysRevA.78.023635, sorensen2001many, jaksch2005cold, goldstein1,santos2000bose}
\begin{IEEEeqnarray}{l}\label{hamilt1}
H=\int{}{d\mathbf{r}}
\sum_{j=a,b}
\wh{\psi}_{j}^{\dagger}(\mathbf{r})
\left(
-\frac{\hbar^{2}}{2m}\nabla^{2}+
V(\mathbf{r})
-\mu_{j}
\right)
\wh{\psi}_{j}(\mathbf{r})
\nn\\
+
\frac{1}{2}
\int{}{d\mathbf{r}}
\sum_{j=a,b}
\wh{\psi}_{j}^{\dagger}(\mathbf{r})
\left(
\sum_{j'=a,b}
g_{jj'}
\wh{\psi}_{j'}^{\dagger}(\mathbf{r})
\wh{\psi}_{j'}(\mathbf{r})
\right)
\wh{\psi}_{j}(\mathbf{r})
\nn\\
+
\frac{\Omega_{R}}{2}
\int{}{d\mathbf{r}}
\left(
e^{i\Delta{t}}
\wh{\psi}_{a}^{\dagger}(\mathbf{r})
\wh{\psi}_{b}(\mathbf{r})
+
e^{-i\Delta{t}}
\wh{\psi}_{b}^{\dagger}(\mathbf{r})
\wh{\psi}_{a}(\mathbf{r})
\right)
.
\end{IEEEeqnarray}
Here $\wh{\psi}_{j}(\mathbf{r})$ is the field operator, which obeys Bose-Einstein statistics and annihilates a particle characterized by the mass $m$, location $\mathbf{r}$, and the internal state $j=a(b)$ for a particle in the excited
(ground) state. The chemical potential for a particle occupying the internal state $j$ is given by $\mu_{j}$. The lattice potential is assumed to have a cubic form, $V(\mathbf{r})=V_{0}\sum_{i=1}^{3}\sin^{2}(k_{L} r_{i})$, and is given in terms of the lattice vector $k_{L}=\pi/a_{L}$, where $a_{L}$ is the lattice constant. The interaction between particles occupying the internal states $j$ and $j'$ are given by the coupling constants $g_{jj'}$. The laser field that drives the condensate from the ground state $|b\rangle$ to the first excited state $|a\rangle$ is characterized by the Rabi frequency $\Omega_{R}$ and detuning $\Delta$ (see Fig.\ref{fig1}.(b)) from the excited state.

In the {tight binding model}, and lowest band approximation, which is valid in the long-wavelength limit, one can expand the bosonic field operators $\wh{\psi}_{j}(\mathbf{r})$ in the Wannier basis $\wh{\psi}_{j}(\mathbf{r})=\sum_{n}b_{nj}w_{j}(\mathbf{r}-\mathbf{r}_{n})$.
{Throughout the paper, we will use the index convention, according to which
the first argument of the index describes the site in an optical lattice, while the second argument corresponds to the internal state within the site.
The expansion of the field operators in the Wannier basis in the driven Bose-Hubbard Hamiltonian {\ref{hamilt1}} directly leads to}
\begin{IEEEeqnarray}{l}\label{hamilt2}
H=-\sum_{j=a,b}
\sum_{
\langle{m,n}\rangle}
J^{jj}_{mn}
\left(
b^{\dagger}_{mj}b_{nj}+
b^{\dagger}_{nj}b_{mj}
\right)
\\\nn
-
\sum_{j=a,b}
\mu_{j}
\sum_{n}
b^{\dagger}_{nj}b_{nj}
+
\sum_{j,j'=a,b}
\frac{U_{jj'}}{2}
\sum_{
n}
b^{\dagger}_{nj}
b^{\dagger}_{nj'}
b_{nj'}
b_{nj}
\\\nn
+
\frac{\Omega_{R}}{2}
\sum_{n}
\left(
e^{i\Delta{t}}
b^{\dagger}_{na}
b_{nb}
+
e^{-i\Delta{t}}
b^{\dagger}_{nb}
b_{na}
\right)
.
\end{IEEEeqnarray}
Here the hopping integral is
\begin{equation}
J^{ij}_{mn}=
-\int
{d\mathbf{r}}
w^{*}_{i}(\mathbf{r}-\mathbf{r}_{m})
\left[
-\frac{\hbar^{2}}{2m}\nabla^{2}+
V(\mathbf{r})
\right]
w_{j}(\mathbf{r}-\mathbf{r}_{n})
,
\end{equation}
and the on-site interaction is
\begin{IEEEeqnarray}{l}
U_{jj'}=
g_{jj'}
\int
{d\mathbf{r}}
w^{*}_{j}(\mathbf{r})
w^{*}_{j'}(\mathbf{r})
w_{j'}(\mathbf{r})
w_{j}(\mathbf{r})
.
\end{IEEEeqnarray}
In order to transform the {\BH} Hamiltonian into the $k$-space we introduce the Fourier transform of the creation and annihilation operators,
\begin{IEEEeqnarray}{l}
b_{nj}=\frac{1}{\sqrt{N_{L}}}
\sum_{k}
\exp{[-i\mathbf{k}\mathbf{r}_{n}]}
a_{kj}
,
\end{IEEEeqnarray}
where $N_{L}$ is number of lattice cites. Then we linearize the Fourier-transformed {\BH} Hamiltonian Eq.(\ref{hamilt2}) by expanding the creation and annihilation operators near their average values,
$a_{kj}=\langle{a_{0j}}\rangle+(a_{kj}-\langle{a_{0j}}\rangle)$. Here $\langle{a_{0j}}\rangle$ is the average value of the  annihilation operator, which is given in terms of number of the particles occupying zero momentum state $N_{0j}$, $\langle{a_{0j}}\rangle=\sqrt{N_{0j}}$. The coupling between particles occupying the internal states $j=\{a,b\}$ is described by the matrix
\begin{eqnarray}\label{intmatrix1}
\left(
\begin{array}{cc}
 n_{a}U_{aa} & \sqrt{n_{a}n_{b}}U_{ab} \\
 \sqrt{n_{a}n_{b}}U_{ba} & n_{b}U_{bb}
\end{array}
\right)\equiv{}
\left(
\begin{array}{cc}
 u_{a} & s_{a} \\
 s_{b} & u_{b}
\end{array}
\right)
.
\end{eqnarray}
Here $n_{j}=N_{0j}/N_{L}$ are the average filling factors of the particles occupying the internal state $j$ and momentum $k=0$. { In this work we will consider the simplified case, $u=n_{a}U_{aa}=n_{b}U_{bb}$ and
$s=\sqrt{n_{a}n_{b}}U_{ab}= \sqrt{n_{a}n_{b}}U_{ba}$. However, the main
physical results concerning the {\Rbg} spectrum of a driven two-component BEC and the characteristic rate of Landau damping of the collective modes in the symmetric case match the results obtained in the general case Eq.(\ref{intmatrix1}).
In the special case $u=s$ the matrix Eq.(\ref{intmatrix1}) simplifies to the fully symmetric Manakov model, which very closely describes a two-component BEC of $^{87}\text{Rb}$.}

\begin{figure}[t!]\label{f1}
\begin{center}
\vspace{-2.5cm}
\includegraphics[angle=0, width=1.0\columnwidth]
{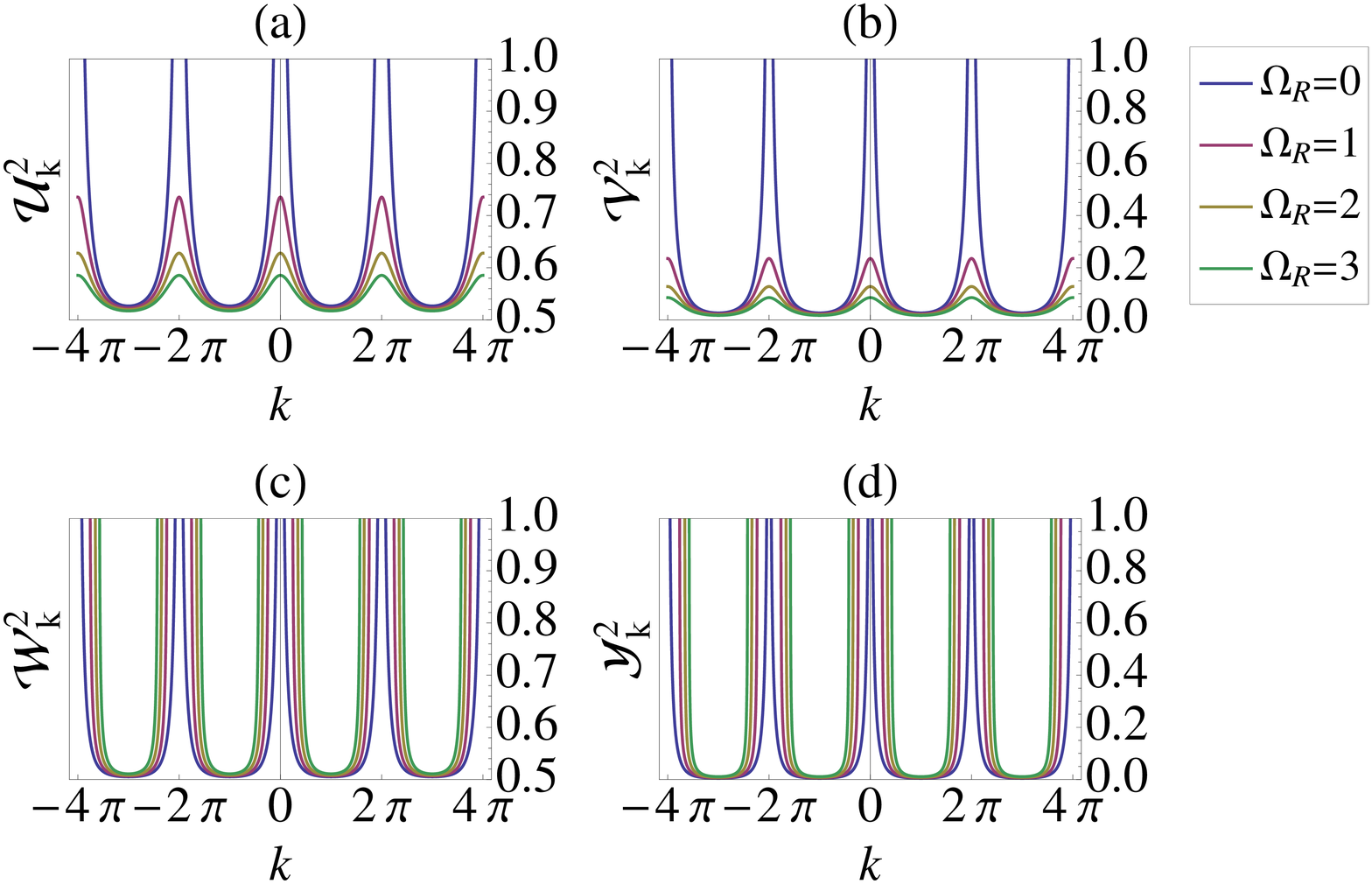}
\end{center}
\vspace{-2.5cm}
 \caption{{\it{\Rbg} amplitudes for a two-component Bose-Einstein condensate  in an optical lattice driven by a microwave field}.
The standard {\Bg} amplitudes have poles at the location of the roots of {\Bg} spectrum. The laser field, which drives the condensate, creates a gap in the spectrum of elementary excitations. As a result, {\Rbg} amplitudes (a) $\mathcal{U}^{2}_{k}$ and (b) $\mathcal{V}_{k}^{2}$ are finite for all values of momenta $k$, given in the units of the lattice constant $a_{L}$. For a two-component condensate, one obtains additional  {\Rbg} amplitudes,
(c) $\mathcal{W}^{2}_{k}$ and (d) $\mathcal{Y}_{k}^{2}$, which are absent for a scalar BEC. These amplitudes ensure Bose-Einstein commutation relation for the quasi-particle creation and annihilation operators.}
\label{ampiltudes1}
\end{figure}

The linearized {\BH} Hamiltonian can be diagonalized via the
generalized {\Bg} transformation. This transformation introduces the quasi-particle creation and annihilation operators, according to $\widehat{a}_{kj}=
\mathcal{U}_{k}\widehat{\alpha}_{ka}+
\mathcal{V}^{*}_{k}\widehat{\alpha}_{-ka}^{\dagger}+
\mathcal{W}_{k}\widehat{\beta}_{kb}+
\mathcal{Y}^{*}_{k}\widehat{\beta}_{-kb}^{\dagger}$.
Here $\widehat{\alpha}_{k,a}$ and $\widehat{\beta}_{k,b}$ are the
quasi-particle annihilation operators in the excited $j=a$ (ground $j=b$). We impose the Bose-Einstein commutation relation for these quasi-particle operators, $[\alpha_{kj}, \alpha^{\dagger}_{-k,j'} ]=\delta_{jj'}$, and $[\beta_{kj}, \beta^{\dagger}_{-k,j'} ]=\delta_{jj'}$. This leads to a constraint on the
{\Rbg} amplitudes,  $\mathcal{U}_{k}^2-\mathcal{V}_{k}^2+\mathcal{W}^{2}_{k}- \mathcal{Y}^{2}_{k}=1$.

In the quasiparticle basis the {\BH} Hamiltonian is diagonal and is given by,
\begin{eqnarray}
H_{\text{eff}}=
\frac{1}{2}\sum_{k}
E_{a}(k)
\widehat{\alpha}_{a,k}^{\dagger}
\widehat{\alpha}_{a,k}+
\frac{1}{2}
\sum_{k}
E_{b}(k)\widehat{\beta}_{b,k}^{\dagger}
\widehat{\beta}_{b,k}
.
\end{eqnarray}
{\Rbg} spectrum of elementary excitations is obtained from the condition
$\det{[M-\mathbf{1}(E/2)]}=0$, where the matrix $M$ is given by
\begin{IEEEeqnarray}{l}
M=\\\nn
\left(
\begin{array}{cccc}
 t_k+u+\frac{\Delta }{2} & s+\frac{\Omega _R}{2} & u & s \\
 s+\frac{\Omega _R}{2} & t_k+u-\frac{\Delta }{2} & s & u \\
 -u & -s & -t_k-u-\frac{\Delta }{2} & -s-\frac{\Omega _R}{2} \\
 -s & -u & -s-\frac{\Omega _R}{2} & -t_k-u+\frac{\Delta }{2}
\end{array}
\right)
.
\end{IEEEeqnarray}
Here, the tunneling parameter $t_{k}=4J\sin^{2}\left({ka_{L}}/{2}\right)$
is given in terms of tunneling amplitude $J\equiv{}J^{jj}_{mn}$, momentum $k$,
and the lattice constant $a_{L}$. The exact {\Rbg} spectrum of a driven two-component condensate is
\begin{IEEEeqnarray}{l}\label{energy1}
E_{a}(k)=
 \sqrt{4 t_k \left(t_k+2 u\right) +\Delta ^2+\Omega_{R}^2+4 s \Omega_{R} + 4 \sigma}
 ,
\nn\\
E_{b}(k)=
 \sqrt{4 t_k \left(t_k+2 u\right) +\Delta ^2+\Omega_{R}^2+4 s \Omega_{R} - 4 \sigma}
 ,
\end{IEEEeqnarray}
where the parameter $\sigma$ is defined as the positive branch of the square root,
\begin{align}
\sigma^{2}&=
4 s t_k \left(t_k+u\right) \Omega _{R} +s^2 \left(4 t_k^2-\Delta ^2\right)
\\\nn
&+\left(t_k+u\right){}^2 \left(\Delta ^2+\Omega _R^2\right)
.
\end{align}
For the {\Rbg} amplitudes we find a set of symmetries that hold among them,
\begin{align}\label{symmetry1}
\mathcal{V}_{k}^2(E_{a},\sigma)
&=-
\mathcal{U}^{2}_{k}(-E_{a},\sigma)
,\\\nn
 \mathcal{W}^{2}_{k}(E_{b},\sigma)
 &=
 \mathcal{U}^{2}_{k}(E_{b},-\sigma)
 ,\\\nn
\mathcal{Y}^{2}_{k}(E_{b},\sigma)
&=-
\mathcal{U}^{2}_{k}(-E_{b},-\sigma)
.
\end{align}
The new symmetries are the direct generalization of the intrinsic symmetries of the standard {\Bg} amplitudes. Indeed, if one reverses the sign of the energy in the square of the first {\Bg} amplitudes, $v^{2}_{k}(E)$, and
swaps the sign of the whole expression, one arrives at the second {\Bg} amplitude, i.e. $u^{2}_{k}(E)=-v^{2}_{k}(-E)$.
As a result of the symmetries Eq.(\ref{symmetry1}), one can obtain all the Rabi-{\Bg} amplitudes from any one of them, for instance, $\mathcal{U}^{2}_{k}(E_{a},\sigma)$, which is explicitly given by,
%
\begin{align}
\mathcal{U}^{2}_{k}(E_{a},\sigma) &=
\frac{1}{4 E_{a}  \sigma}
\times
\left(
s^2 \left(4 t_k-2 \Delta \right)
\right.
\\\nn
&+
\left.
\left(2 t_k+2 u+E_{a} +\Delta \right) \left(\left(t_k+u\right) \Delta +\sigma \right)
\right.
\\\nn
&+
\left.
2 s \left(2 t_k+u\right) \Omega _R+\left(t_k+u\right) \Omega _R^2
\right)
\end{align}
%
{
We note that in the long-wavelength limit the Rabi-{\Bg} amplitudes $\mathcal{W}_{k}(E_{a})$ and $\mathcal{Y}_{k}(E_{a})$ are purely imaginary.
Therefore, we are left with the real-valued Rabi-{\Bg} amplitudes $\mathcal{U}_{k}(E_{a})$ and $\mathcal{V}_{k}(E_{a})$, which can be considerably simplified in the special case of a resonant drive, i.e. $\Delta=0$,}
\begin{IEEEeqnarray}{l}
\mathcal{U}_{k}(E_{a})=\sqrt{
\frac{E_{a} +2 (t_k+u+s+\Omega _R/2)}
{4 E_{a} }}
,
\\\nn
\mathcal{V}_{k}(E_{a})=-\sqrt{
\frac{-E_{a} +2 (t_k+u+s+\Omega _R/2)}
{4 E_{a} }},
\end{IEEEeqnarray}
where $E_{\alpha}=\sqrt{\left(2 t_k+\Omega _R\right) \left(2 t_k+4 u+4 s+\Omega _R\right)}$. If we introduce $E=E_{\alpha}/2$, we obtain  Landau decay rate of the collective modes in a driven two-component BEC,
\begin{align}\label{landauint1}
\Gamma_{L}&=
-\pi \hbar \omega_{q}
\frac{2\pi}{(2\pi\hbar)^{3}}
\left(
4 \sqrt{N}\frac{g_{jj}}{2}
\frac{\sqrt{\omega_{q} }}{\sqrt{2} \sqrt{u+s}}
\right)^{2}
\frac{1}{q}
\nn\\
&\times
 \beta
\frac{\p}{\p \beta}
\int{}
{dp}\;
\frac{1}{v_{g}}
\frac{p^{2}}{E}
\frac{1}{(e^{\beta  E}-1)}
\left(
\frac{3}{4}\frac{E}{(u+s)}
\right)^{2}
,
\end{align}
where $\omega_{q}$ and $q$ are the frequency and momentum of the collective mode. Now we define the density of the roton gas,
\begin{align}\label{densityrot1}
\rho_{r}&=
\frac{4\pi}{3(2\pi\hbar)^{3}}
\int{}
{dp}\;
p^{2}
\frac{E^{2}}{v_{g}}
\left(-\frac{\p}{\p E}
\frac{1}{(e^{\beta  E}-1)}
\right)
\nn\\
&=
\frac{4\pi}{3(2\pi\hbar)^{3}}
\left(-\beta\frac{\p}{\p \beta}\right)
\int_{E_{0}}^{\infty}
{dE}\;
\frac{p^{2}}{v_{g}^{2}}
\frac{E}{(e^{\beta  E}-1)}
,
\end{align}
where  the group velocity is defined as $v_{g}={\p E(p)}/{\p p}$
and $\beta=1/k_{B}T$, given in terms of the Boltzmann constant $k_{B}$.
For small momenta we find laser-induced roton-like spectra, $E\simeq{}E_{0}+E_{2}p^{2}/2$, where we have introduced the gap $E_{0}=\sqrt{\Omega _R \left(4 u+\Omega _R\right)}/2$ and the curvature $E_{2}=\left(2 u+\Omega _R\right)/
[m^{*} \sqrt{\Omega _R \left(4 u+\Omega _R\right)}]$,
given in terms of the effective mass $m^{*}={1}/(Ja_{L}^{2})$.  Then we immediately obtain Landau damping rate of the two-component BEC in an optical lattice expressed in terms of the density of the laser-induced roton mode,
\begin{equation}\label{landaurot1}
\Gamma_{L}=
\theta(\hbar\omega_{q}-E_{0})
\frac{27\pi}{16}
\hbar\omega_{q}
\frac{\rho_{r}}{\rho(\omega_{q})}
,
\end{equation}
where the collective mode spectral density
$
\rho(\omega_{q})=
q(u+s)^{3}/(g_{jj}^{2}N\omega_{q})$ and the momentum $q=
\sqrt{{2(\hbar\omega_{q}-E_{0})}/{E_{2}}}$.
For a collective mode with energy below the gap, i.e. $\hbar\omega_{q} < E_{0}$,
the Landau damping is zero, $\Gamma_{L} = 0$.
Thus, laser-induced gap in the spectrum protects the collective modes from Landau damping.
Above energy of the gap, the Landau damping rate is proportional to the density of the laser-induced roton gas, which scales at low temperatures as  $\rho_{r}\simeq{}\frac{1}{\beta^{2}}$.
In the special field-free case, i.e. $\Omega_{R}=\Delta=0$, {\Rbg} spectrum reduces to the
conventional {\Bg} spectrum. Thus, we obtain well-known result \cite{pitaevskii1997landau} of the phonon-mediated Landau damping of the collective modes,
\begin{equation}
\Gamma_{L}(\Omega_{R}=0)=
\frac{27\pi}{16}
\hbar\omega_{q}
\frac{\rho_{n}}{\rho}\simeq{}\frac{1}{\beta^{4}}
.
\end{equation}
Here $\rho_{n}$ is the normal density of a phonon gas
$\rho_{n}=2\pi^{2}T^{4}/({45\hbar^{3}c^{5}})$, where $c$ is the speed of sound \cite{landau1980statistical}. Therefore, in the presence of a laser field
damping of the collective modes in the condensate significantly slows down
compared to the laser-free phonon-mediated Landau damping of a scalar BEC.

 {
The damping rate of collective excitations in a driven Bose gas
can be verified via two-photon Bragg spectroscopy as it was done in \cite{PhysRevLett.89.220401}. In the experiment, the quasiparticles in a BEC of $^{87}{\rm Rb}$ atoms were excited by tuning the frequency difference and angle between the Bragg beams applied to the condensate. Immediately after the applied Bragg pulses, the
magnetic trap, which confined the BEC, was rapidly
turned off. Following the free expansion the BEC cloud was imaged via
on-resonance absorption, which allowed the number of scattered atoms as a function of energy and momentum to be extracted. The measurement revealed a significant
suppression of collisions of quasiparticles at low momentum. In both cases, namely, Beliaev damping of collective modes in a laser-free BEC, and Landau damping in a laser-driven condensate, the physical systems are characterized by a critical energy, below which collision of the collective modes and associated damping processes are completely suppressed. We conclude that even though the collisions between quasiparticles in the experiment were described by the Beliaev damping process, we expect that the same holds for Landau damping of collective modes in a laser-driven condensate.}

 {
In conclusion, we have investigated the quantum many-body physics of a driven two-component Bose-Einstein condensate in an optical lattice driven by a microwave field. We derived exact analytical results for the generalized {\Rbg} spectrum and amplitudes of the condensate. We found a gap in the spectrum of elementary excitation in the BEC, which amounts to the effective Rabi frequency of the applied laser field. We discovered symmetries between the elementary excitations of a driven Bose gas, which generalize the underlying symmetries in the standard {\Bg} amplitudes. The gapped spectrum and new symmetries
of elementary excitations in a driven BEC dramatically modify dynamics of collective modes
compared to the laser-free case. Specifically, we found that below the gap energy the collective mode are damping-free. Above the gap energy the damping rate is proportional to the density of the laser-induced roton mode. Thus, the Landau damping rate of the collective modes in a driven condensate is considerably reduced compared
to the phonon-mediated damping processes in a laser-free condensate.}

{\it Acknowledgments}
This material is based in part upon work supported by the US National Science Foundation under grant numbers PHY-1306638, PHY-1207881, and PHY-1520915, and the US Air Force Office of Scientific Research grant number FA9550-14-1-0287.


%

\end{document}